\documentclass{svmult}

\newcommand{\ali}[1]{\begin{align}#1\end{align}}
\def\br{\textbf{r}}
\def\bv{\textbf{v}}
\newcommand{\eref}[1]{(\ref{#1})}

\newcommand{\fref}[1]{Fig.~\ref{#1}}

\def\rr{\mathbf{r}}
\def\kk{\mathbf{k}}

\def\intr{\int d^3\rr\;}

\def\del{\partial}

\usepackage{graphicx}        
\graphicspath{{./figures/}}
\usepackage{placeins}        
\usepackage{tabularx}        
\usepackage{multicol}        
\usepackage[bottom]{footmisc}

\usepackage{soul}
\usepackage{newtxtext}       %
\usepackage{newtxmath}       

\usepackage[bookmarks=false,colorlinks=true,linkcolor=blue,citecolor=blue,urlcolor=blue]{hyperref}
\providecommand{\doi}[1]{\href{https://doi.org/#1}{\nolinkurl{https://doi.org/#1}}}

\makeindex             

\begin{document}

\title*{Turbulence in Quantum Gases: Vortices, Waves, and Cascades}
%
\author{
Ashton S. Bradley
\and
Tyler W. Neely
\and
Xiaoquan Yu
\and
Brian P. Anderson
}

\institute{
Ashton S. Bradley \at
Department of Physics, University of Otago, Dunedin, New Zealand\\
\email{ashton.bradley@otago.ac.nz}
\and
Tyler W. Neely \at
School of Mathematics and Physics, University of Queensland, Brisbane, Australia
\and
Xiaoquan Yu
\at
Graduate School of China Academy of Engineering Physics, Beijing, China
\and
Brian P. Anderson
\at
Wyant College of Optical Sciences, University of Arizona, Tucson, USA
}
\maketitle
\abstract
{
We review turbulence in ultracold quantum gases, using the scalar contact-interaction Bose-Einstein condensate as the reference system for quantized circulation, compressibility, vortices, sound, and cascades. We focus on the quantitative diagnostics that connect helium and classical phenomenology to microscopic wave-function dynamics: incompressible and compressible kinetic-energy spectra, wave-occupation spectra, spectral fluxes, vortex-resolved correlations, and velocity statistics. These diagnostics distinguish equilibrium vortex organization, decaying turbulent relaxation, forced cascade dynamics, and weak-wave turbulence, and show why power laws alone are insufficient evidence for a cascade. We survey experiments on two-dimensional Onsager clustering, three-dimensional vortex-line turbulence, box-trap wave cascades, engineered dissipation, and turbulent equations of state. We close by briefly placing the contact-interaction scalar superfluid system in a broader landscape of nonlocal, multicomponent, fermionic, and driven-dissipative quantum fluids, where turbulence concepts can be tested for universality.
}


\newpage

\section{Foundations of quantum-gas turbulence}

\subsection{History and phenomenology of superfluid turbulence}
\label{sec:intro}
The observation of Bose-Einstein condensation (BEC) in dilute atomic gases in 1995~\cite{anderson_observation_1995} realized a long-predicted phase of matter with macroscopic quantum coherence, the matter-wave counterpart of optical coherence in laser light~\cite{cornell_nobel_2002,ketterle_nobel_2002}. Ultracold gases are clean, dilute, and highly controllable quantum fluids, routinely prepared close to their many-body ground state. They therefore provide a setting in which the defining ingredients of superfluid turbulence--phase coherence, quantized circulation, vortices, compressibility, and sound--can be connected directly to microscopic wave-function dynamics.

In a classical viscous fluid, turbulence is often controlled by the Reynolds number $Re=UL/\nu$, where $U$ is a characteristic velocity, $L$ a characteristic length, and $\nu$ the kinematic viscosity. A large Reynolds number expresses the dominance of inertial transport over viscous damping, and underlies the familiar cascade phenomenology of classical turbulence. Quantum gases do not map directly onto this definition, because an ideal superfluid has no classical viscosity; the useful analogy must instead be built from quantized circulation, compressibility, and sound emission and interaction. 

The subject inherits much of its language from helium. Onsager recognized that the single-valued superfluid phase makes circulation around loops enclosing vortices quantized, with quantum scale set by $h/m$, Planck's constant over the atomic mass; he also connected the dynamics of vortices carrying this circulation to point-vortex statistical mechanics~\cite{onsager_statistical_1949}; Feynman described superfluid turbulence as a tangle of vortex lines, each carrying the same circulation quantum~\cite{feynman_application_1955}; and Vinen developed the phenomenology of vortex-line turbulence and mutual friction~\cite{vinen_mutual_1957,vinen_introduction_2006}. Atomic gases add microscopic control and direct imaging to this lineage. Vortices and vortex lattices were created soon after the first condensates~\cite{matthews_vortices_1999,madison_vortex_2000,abo-shaeer_observation_2001}, and later experiments have used stirring, box confinement, and in situ imaging to probe turbulent transport in increasingly programmable quantum fluids~\cite{neely_observation_2010,navon_emergence_2016,gauthier_giant_2019,johnstone_evolution_2019,zhao_kolmogorov_2025}.

The first distinction from typical classical viscous fluids is topological. In a dilute-gas BEC, the quantum fluid is described by a macroscopic order parameter whose phase encodes long-range coherence, and phase gradients determine the superfluid velocity. Single-valuedness of the wave function restricts phase winding around closed contours to integer multiples of $2\pi$\footnote{Onsager famously pointed this out in a footnote~\cite{onsager_statistical_1949}.}. Circulation is therefore quantized, and vorticity is localized on vortex lines in three dimensions or point vortices in effectively two-dimensional condensates, embedded in an otherwise irrotational flow. Classical fluids instead allow continuously distributed vorticity and circulation~\cite{vinen_introduction_2006,nemirovskii_quantum_2013}.

Compressibility supplies the second essential distinction. The size of a quantum-vortex core is set by the healing length, where kinetic energy from phase winding is balanced by the interaction energy cost of density depletion. The same interaction scale determines the sound speed, so vortices and density waves are not independent embellishments of the theory. This is the physical reason for first developing the Gross-Pitaevskii equation as a hydrodynamic theory, and then using the theory to identify turbulent regimes.

It is important to separate the classical-field and quantum phenomena that occur in a turbulent superfluid. We use \emph{quantum turbulence} in the strict sense of turbulent dynamics involving topological excitations of the order parameter, principally quantized vortices and, where their phase topology is essential, solitonic defects. The Bogoliubov phonon branch is quantum in origin: it is the Goldstone branch of the condensate, determines the superfluid sound speed, and underlies the Landau criterion. However, wave turbulence formed by highly occupied, smooth Gross--Pitaevskii modes is a classical-field phenomenon unless its dynamics require nonclassical correlations or operator physics beyond the coherent mean field. Superfluid turbulence can contain classical-field waves, topological quantum turbulence, or a mixed regime in which vortices and waves exchange energy. The interplay between these sectors is a central theme of this review.

Recent experiments have shifted the field from the observation of vortices and turbulent-looking states toward quantitative tests of transport. Box traps, programmable optical potentials, vortex-sign detection, engineered loss, momentum-resolved measurements, and velocity statistics now make it possible to compare forcing, dissipation, momentum distributions, fluxes, vortex statistics, and wave occupations within the same quantum fluid. This is the central opportunity of atomic-gas turbulence: to turn cascade phenomenology into a controlled microscopic nonequilibrium process in which Gross-Pitaevskii and wave-kinetic theory can be tested against resolved experiments.

\subsection{Gross-Pitaevskii equation}
\label{sec:theory}
Cold weakly interacting BECs are well described by the Gross-Pitaevskii equation~\cite{pitaevskii_bose-einstein_2003}. For identical bosons with field operator $[\hat\psi(\br,t),\hat\psi^\dagger(\br',t)]=\delta(\br-\br')$, the near-zero-temperature condensate may be described by the coherent mean field $\psi(\br,t)\equiv\langle\hat\psi(\br,t)\rangle$. This order parameter evolves according to the Gross-Pitaevskii equation (GPE)
\begin{align}
	i\hbar\frac{\partial \psi}{\partial t}=\left(-\frac{\hbar^2\nabla^2}{2m}+U(\br,t)+\varg|\psi|^2\right)\psi,
\end{align}
where $U$ is the potential energy, $m$ is the atomic mass, and $\varg=4\pi\hbar^2 a/m$ is the three-dimensional two-body contact interaction strength for $S$-wave scattering length $a$. There are many routes to the GPE, all relying on quantum coherence and giving slightly different interpretations of $\psi(\br,t)$~\cite{blakie_dynamics_2008,proukakis_finite-temperature_2008}. For the present purpose its importance is that a single nonlinear wave equation contains vortices, sound, density depletion, and their mutual conversion.

The GPE conserves the total particle number $N$ and energy $E$, defined as\footnote{For brevity, in what follows we will suppress space-time arguments unless there is risk of ambiguity.}
\ali{\label{Ngp}
N&\equiv\intr |\psi|^2,\\
\label{Hgp}
E&\equiv\intr \left[\frac{\hbar^2}{2m}|\nabla\psi|^2+U|\psi|^2+\frac{\varg}{2}|\psi|^4\right].
}
For a stationary constrained GPE solution, the chemical potential $\mu(N)$ is defined by
\ali{
\mu(N)&=\frac{\partial E}{\partial N}=\frac{1}{N}\intr \left[\frac{\hbar^2}{2m}|\nabla\psi|^2+U|\psi|^2+\varg |\psi|^4\right].
}
For the homogeneous ground state of a system confined to a hard-wall box of volume $V$, $|\psi|^2\to n_0=N/V$ and $\mu=\varg n_0$. The homogeneous system also suggests natural units of length and time set by the interaction energy, and defined via the relationship
$\mu= \varg n_0 \equiv \hbar^2/(m\xi^2)\equiv mc^2.$
This serves to introduce the healing length and speed of sound,
\ali{\xi&=\frac{\hbar}{\sqrt{m\varg n_0}}=\frac{\hbar}{mc},\quad\quad c=\sqrt{\frac{\varg n_0}{m}}, }
key physical parameters of a compressible quantum fluid that play a central role in quantum fluid dynamics and quantum turbulence.

The transition from nonlinear wave equation to fluid mechanics is made by the Madelung representation $\psi=\sqrt{n}e^{i\Theta}$,
where $n(\rr,t)=|\psi(\rr,t)|^2$ is the particle density and $\Theta(\rr,t)$ is the phase of the quantum field $\psi$, and plays the role of a velocity potential for the quantum fluid. Wherever $n$ is finite this transformation defines the superfluid velocity field
\ali{\label{velocity}
\mathbf{v}(\rr,t)&\equiv\frac{\hbar}{m}\nabla\Theta(\rr,t)}
and makes the phase winding of $\psi$ a circulation constraint. Around any simple closed contour $C$,
\ali{\label{qcirc}
\Gamma&\equiv\oint_C \bv\cdot \mathbf{d}\br = \frac{h}{m}\sum_i q_i,
}
with integer charges $q_i\in\{\dots,-1,0,1,2,\dots\}$ for each enclosed vortex\footnote{Only charges $\pm 1$ are dynamically stable, and higher charge excitations rapidly fragment into multiple unit charge vortices.}. The existence of a velocity potential for the superfluid means that formally the flow is curl-free wherever the density is nonzero. In an effectively two-dimensional condensate, the scalar vorticity normal to the plane is isolated to point-vortex cores,
\ali{
(\nabla\times\bv)_z&=\frac{h}{m}\sum_i q_i\delta^{(2)}(\br-\br_i).
}
In three dimensions the vorticity is located at vortex filaments and directed along their local tangents. The single-valued phase of the macroscopic order parameter is thus the origin of the topological nature of quantum vortices.

Using the Madelung transformation, the Hamiltonian becomes
\ali{\label{hydroH}
H=\underbrace{\intr \frac{m}{2}n|\bv|^2}_{\text{hydrodynamic}}+\underbrace{\intr\frac{\hbar^2}{2m}
|\nabla\sqrt{n}|^2}_{\text{quantum pressure}}+\underbrace{\intr U(\rr,t)n}_{\text{potential}}+\underbrace{\intr\frac{\varg}{2}n^2}_{\text{interaction}},
}
with a direct physical interpretation. The hydrodynamic kinetic energy is the quantity most often analysed in quantum turbulence, because it contains the vortex and acoustic motion that can form scale-local cascades. A spectral decomposition will be introduced below once the vortex core has been identified as a finite-density-depletion structure of a compressible quantum fluid. The remaining energy terms identify the other physical scales: quantum pressure resolves density gradients and vortex cores, the potential energy encodes confinement and forcing, and the interaction energy sets compressibility and the energy cost of density modulation.

\subsection{Inviscid hydrodynamics, pressure, and incompressibility}
The Madelung form of the GPE provides the bridge to classical hydrodynamics. Away from vortex cores and sharp density gradients it reduces to an inviscid compressible Euler fluid, while the quantum pressure and quantum stress terms control the breakdown of this approximation. This formulation has been used to describe flow past obstacles and the onset of vortex shedding in quantum fluids~\cite{frisch_transition_1992,winiecki_pressure_1999,winiecki_vortex_2000}.

We first introduce the particle current
$\mathbf{J}(\rr,t)\equiv \left(\psi\nabla\psi^*-\psi^*\nabla\psi\right)i\hbar/(2m)=n \mathbf{v}$. The GPE can then be recast as a continuity equation and a superfluid Euler equation for the velocity field, 
\ali{
\del_t n(\rr)+\nabla\cdot\mathbf{J}(\rr)&=0,\\
\del_t \bv(\rr)+(\bv\cdot\nabla)\bv&=-\frac{1}{m}\nabla U_{\rm eff}(\rr),
}
where the effective potential is
\ali{\label{veff}
U_\textrm{eff}(\rr)&\equiv  U+ \varg n-\frac{\hbar^2}{2m}\frac{\nabla^2\sqrt{n}}{\sqrt{n}} .
}
The final term is the quantum potential. It is negligible for smooth density profiles on scales large compared with $\xi$, but it is essential near vortex cores, solitonic density notches, and short-wavelength density fluctuations associated with shocks.

The Madelung velocity formulation is not, by itself, a globally regular dynamical description of vortical flows. The velocity field diverges at a vortex core and the phase is undefined where $n\to 0$, allowing the phase to acquire a topological winding. Correctly accounting for the multivalued phase near vortex cores is therefore a central subtlety of hydrodynamic descriptions~\cite{dos_santos_hydrodynamics_2016}. A well-defined dynamical statement can instead be made in terms of the particle current, which vanishes at vortex cores. Summing over repeated indices, the equations of motion for particle density and current are
\ali{
\del_t n+\del_k J_k&=0,\\
m\del_t J_k+\del_j T_{j k}&=-n\del_k U,
}
where the momentum-flux tensor~\cite{winiecki_vortex_2000} is
\ali{\label{fdtensor}
T_{jk}&\equiv\frac{\hbar^2}{4m}\left(\del_j\psi^*\del_k\psi-\psi^*\del_{jk}\psi+\text{c.c}\right)+\frac{\varg}{2}|\psi|^4\delta_{jk}.
}

In Madelung form this flux tensor may be written directly as
\ali{\label{fdtensor_madelung}
T_{jk}&=m n\varv_j \varv_k+\frac{\varg}{2}n^2\delta_{jk}
-\frac{\hbar^2}{4m}n\del_j\del_k\ln n .
}
The first term is the convective momentum flux, the second generates the bulk hydrostatic pressure, and the final quantum-stress contribution describes kinetic stresses from density gradients. This last term is physically significant in regions of low density and high density curvature, such as vortex cores and short-wavelength density waves.

For a fluid parcel at rest, we define the scalar pressure as the isotropic part of the non-convective momentum flux. In $D$ dimensions this is
\ali{\label{pres}
P &=\frac{\varg n^2}{2}-\frac{\hbar^2}{4m}\frac{n}{D}\del_j\del_j\ln n,
}
Here $\varg n^2/2$ is the interaction or equation-of-state pressure, while the second term is the isotropic trace contribution of the quantum-stress tensor; the factor $D^{-1}$ averages its normal components. The scalar pressure vanishes at a vortex centre, even though the quantum-stress tensor remains finite and anisotropic in the core limit. In a system with uniform particle density $n_0$, the pressure is
$P_0=\tfrac{1}{2}\varg n_0^2$. A uniform weakly interacting gas hence has isentropic compressibility $\beta_s\equiv n^{-1}\partial n/\partial P=1/(\varg n_0^2)=1/(2P_0)$, and a large or strongly interacting condensate approaches approximately incompressible flow at sufficiently low Mach number $M\equiv \varv/c$. For a flow with speed $\varv$ we can estimate the pressure variation in a uniform system using the Bernoulli balance scaling in the flux tensor, $\varg n^2\sim mn\varv^2$. Hence $\Delta n/n\sim\beta_s\Delta P\sim (\varv/c)^2=M^2\ll 1$ and the fluid is incompressible at leading order in $M$. This low-Mach, weakly compressible limit is precisely the regime in which the incompressible kinetic-energy spectrum becomes a meaningful diagnostic of vortex-dominated motion. Deviations from this limit appear as compressible kinetic energy and density-wave occupation, while a small core size $\xi$ compared with the typical vortex separation and system size permits point-vortex or vortex-line descriptions.

\subsection{Phonons and superfluidity}
The compressible sector of the GPE is formed by Bogoliubov phonons. Linearising the hydrodynamic equations on a stationary homogeneous background $n=n_0+\delta n$, to first order in $\bv$ and $\delta n$, gives the density-wave equation
\begin{align}
    \frac{1}{c^2}\frac{\partial^2 \delta n}{\partial t^2}&=\nabla^2\delta n-\frac{\xi^2}{4}\nabla^4\delta n,
\end{align}
where the dispersive term is the quantum-pressure correction written using $\xi=\hbar/\sqrt{m\varg n_0}=\hbar/(mc)$. Small-amplitude plane waves of the form $\delta n\propto \textrm{Re}\{ e^{i(\kk\cdot\br-\omega(k)t)}\}$ satisfy the Bogoliubov dispersion relation
\ali{\label{phonon}
\hbar\omega(k)&=\sqrt{\frac{\hbar^2k^2}{2m}\left(\frac{\hbar^2k^2}{2m}+2\varg n_0\right)}.
}
Phonons in the homogeneous BEC thus have phase velocity $\varv_p=\omega/k=c\sqrt{1+(k\xi/2)^2}$. Long-wavelength excitations are acoustic, while short-wavelength excitations approach free-particle behavior $\hbar\omega(k)\to \mu+\hbar^2k^2/2m$ when $k\xi \gg 1$. The crossover near $k\xi\sim 1$ is where compressible density waves most strongly feel the microscopic structure of the condensate.

The same spectrum sets the ideal Landau criterion. In a homogeneous weakly interacting condensate, the critical velocity obtained from $\min_k\omega(k)/k$ is the sound speed $c$. This criterion identifies the energetic onset of compressible excitations, but real superfluid breakdown in finite geometries is usually mediated by density inhomogeneity, obstacle geometry, and vortex-core formation. Smooth imposed flow can therefore convert into phonon emission, vortex nucleation, and vortex shedding at velocities below $c$; the corresponding vortex processes and experiments are discussed in Sec.~\ref{sec:vortex_processes}. This link between vortices, sound, compressibility, and dissipation is the reason that phonon spectra and vortex dynamics cannot be separated in a turbulent Gross-Pitaevskii fluid.

\section{Turbulent degrees of freedom and regimes}
The contact-interaction Gross-Pitaevskii fluid supports two coupled turbulent sectors: quantum vortex motion, and classical-field wave motion usually associated with smooth density and phase fluctuations~\footnote{The strongly forced regime is an important exception we will encounter shortly.}. Which sector dominates depends on dimensionality, excitation energy, compressibility, forcing scale, dissipation scale, and temperature. These variables provide a guide for interpreting simulations and experiments. Section~\ref{sec:intro} established the hydrodynamic variables and stresses; the present section organizes the resulting vortex, wave, and mixed regimes; the following section describes experimental approaches to creating and detecting different regimes.

\subsection{Regimes: vortices, waves, and mixed turbulence}
In three dimensions, the simplest regime sequence is controlled by excitation energy, setting the relative weight of waves and vortex lines. In the weak-wave regime, classical-field turbulence involves weakly nonlinear mixing of Bogoliubov modes without topological excitations. Near the vortex-nucleation threshold, density waves, phase fluctuations, and sparse vortices coexist, producing mixed classical-wave and quantum-vortex turbulence. At higher excitation energy, the flow is dominated by a dense tangle of quantized vortex lines, reconnections, Kelvin-wave excitations, sound emission, and reduced long-range phase coherence. We refer to this vortex-dominated, strongly forced regime as \emph{strong vortex turbulence}.

Within the strong three-dimensional regime one should further distinguish Vinen/ultraquantum turbulence from quasi-classical Kolmogorov turbulence~\cite{barenghi_types_2023}. Vinen/ultraquantum turbulence is dominated by a random vortex tangle, vortex-line density decay, reconnections, Kelvin-wave excitations, and sound radiation, without a large-scale polarized flow. Quasi-classical turbulence additionally contains bundles or polarization of vortex lines that support a Kolmogorov-like incompressible cascade over scales larger than the intervortex spacing. Atomic gases can probe both limits, but finite system size, density inhomogeneity, and imaging resolution make the distinction more delicate than in an ideal homogeneous tangle.

Two-dimensional turbulence is organized differently because plane-confined vortices are point-like and interact logarithmically over long distances. Opposite-sign vortices form translating dipoles at low energy, while unbound neutral mixtures form vortex plasmas. At high point-vortex energy, same-sign vortices cluster into Onsager states with large-scale circulation and negative absolute temperature~\cite{onsager_statistical_1949,reeves_inverse_2013, billam_onsager-kraichnan_2014,reeves_turbulent_2022}. These states provide a quantum route to inverse-transfer phenomenology: energy can move to larger length scales through vortex clustering rather than being transported only to small dissipative scales.

Dimensionality is only one parameter setting the regime. Forcing determines whether energy is injected mainly as sound, isolated vortex dipoles, rotating same-sign vortices, vortex rings, surface waves, or mixed excitations. Compressibility controls the ease with which vortex motion radiates sound. Dissipation may be set by trap loss, ultraviolet evaporation in a finite-depth box, thermal damping, vortex-antivortex annihilation, or phonon emission. Onsager noted that the circulation quantum $h/m$ has the same dimensions as a classical kinematic viscosity, motivating a heuristic link between vortex-mediated superfluid breakdown and an effective quantum viscosity~\cite{onsager_introductory_1953}. Two-dimensional vortex-shedding calculations and the associated Strouhal oscillations support an effective superfluid Reynolds-number description in that setting~\cite{reeves_identifying_2015,christenhusz_emergent_2025}, but the question of a universal constitutive viscosity for quantum fluids remains open. More generally, vortex formation occurs when the local flow violates the relevant Landau or geometry-dependent critical-flow condition; subsequent vortex production, motion, annihilation, and loss can then provide dissipation-like channels. Such analogies are useful once the active vortex and wave degrees of freedom have been identified by diagnostics that separate spectra, fluxes, and vortex correlations.

\subsection{Quantum vortices, creation, and elementary processes}
\label{sec:vortex_processes}
\label{sec:vortex_generation}
\subsubsection{Vortex structure and vortex pairs}
\begin{figure}[!t]
\centering
	\includegraphics[width=\textwidth]{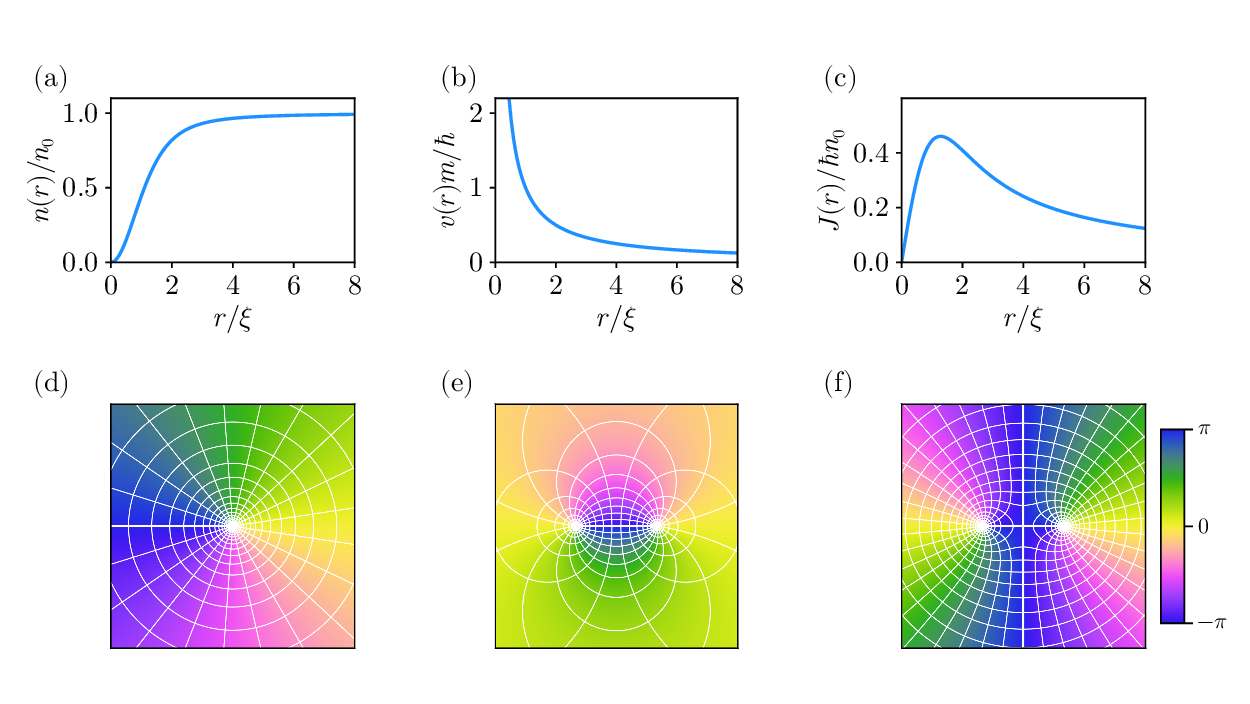}
	\caption{Vortex properties for the GPE: (a) numerically computed stationary radial GPE density profile of a singly charged vortex, (b) velocity, (c) current, and quantum phase (colour) with streamlines for (d) a single vortex, (e) a vortex--antivortex dipole with vanishing net phase winding in the far field, and (f) a same-sign vortex pair with a stagnation point between the vortices and $4\pi$ phase winding in the far field}
    \label{vort_prop}
\end{figure}

The elementary vortical excitation of the GPE is the singly charged quantum vortex. Its existence depends on both phase topology and compressibility. The phase winding fixes the circulation, while compressibility allows the density to fall to zero over a finite core of size $\xi$, avoiding a singular kinetic-energy density at the vortex centre. The density, velocity, current, phase, and streamlines of isolated and paired vortices are shown in \fref{vort_prop}. At the vortex centre the density and pressure \eref{pres} vanish, regularizing the divergent velocity field. In dimensionless form, $m\xi |\mathbf{v}|/\hbar=\xi/r$, so the velocity scale near the core is set by $\hbar/(m\xi)$, while the particle current vanishes linearly as the density depletion regularizes the divergent velocity.

Dipoles and same-sign pairs place vortex kinetic energy at different scales. A vortex-antivortex dipole has strongly suppressed far-field velocity because the two circulations cancel; its energy is therefore localized near the dipole separation and the vortex cores, and it contributes mainly at wavenumbers $k\gtrsim d^{-1}$ for dipole size $d$. A same-sign vortex pair has the far field of a multiply charged vortex. Its circulation reinforces at large distances, producing large-scale flow and enhanced low-$k$ vortex energy. In many-vortex gases this distinction between opposite-sign dipoles and same-sign pairs is the microscopic origin of the spectral contrast between small-scale vortex-dipole energy, uncorrelated vortex-plasma energy, and large-scale Onsager clustering.

\subsubsection{Basic vortex creation mechanisms}
Superfluid breakdown under imposed flow is one important source of these vortical degrees of freedom. Near an obstacle the density is depleted, the local sound speed $c(\br)=\sqrt{\varg n(\br)/m}$ is reduced, and the flow is accelerated around curved boundaries. Vortex emission is therefore triggered by the maximum local Mach number rather than by the imposed far-field speed alone~\cite{frisch_transition_1992}. Above threshold, a moving obstacle can shed vortex dipoles~\cite{frisch_transition_1992}; increasing the obstacle speed or size creates higher-energy flows including charge-2 vortex streets, periodic vortex-cluster shedding, and oblique soliton trains~\cite{sasaki_benard--von_2010}. In the oblate obstacle experiment of Neely \textit{et al.}, vortex dipoles were generated at imposed speeds above $\sim 0.1c$~\cite{neely_observation_2010}; this value is protocol-specific because the threshold depends on obstacle geometry, trap inhomogeneity, and the local vortex-nucleation condition.

\begin{figure}[!t]
\centering
\includegraphics[width=\textwidth]{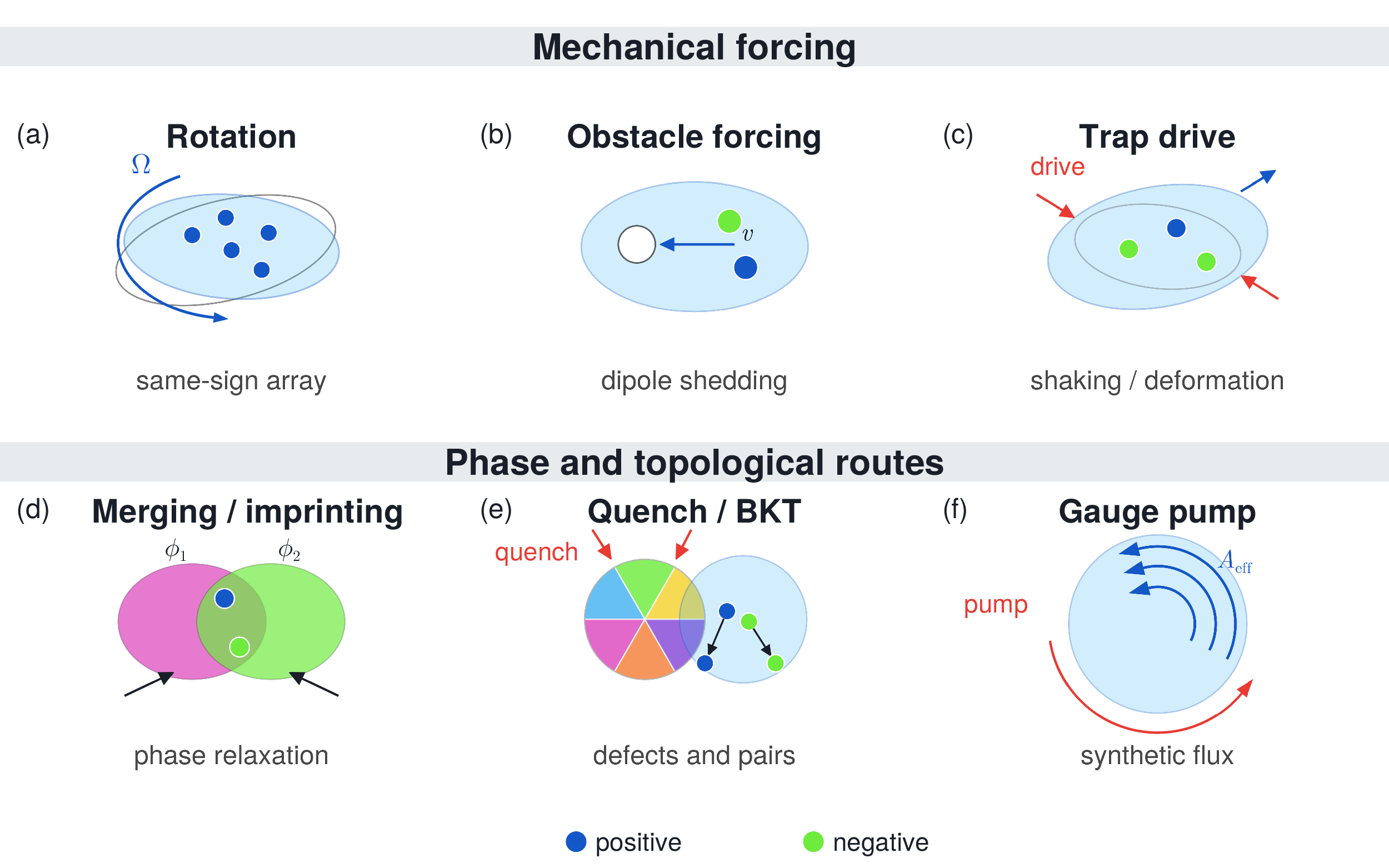}
	\caption{Schematic mechanisms for vortex formation in quantum gases. Mechanical forcing creates circulation by rotating the gas [panel (a)]~\cite{matthews_vortices_1999,madison_vortex_2000}, by shedding, splitting, or pinning vortices with localized optical potentials [panel (b)]~\cite{raman_evidence_1999,neely_observation_2010,samson_deterministic_2016}, or by driving the trap [panel (c)]~\cite{henn_emergence_2009,navon_emergence_2016}; panel (c) represents time-dependent shaking or deformation of the confinement, which injects energy until shape instabilities nucleate vortex cores. Phase and topological routes create vortices when condensates with distinct phase are merged or directly imprinted [panel (d)]~\cite{scherer_vortex_2007,hernandez-rajkov_connecting_2024,del_pace_imprinting_2022}, when phase domains freeze out~\cite{haljan_vortices_2003,weiler_spontaneous_2008} or BKT pairs unbind [panel (e)]~\cite{hadzibabic_berezinskii-kosterlitz-thouless_2006}, or when synthetic gauge fields and topological pumping protocols impose circulation [panel (f)]~\cite{lin_synthetic_2009,leanhardt_imprinting_2002,moon_thermal_2015}}
	    \label{fig:vmech}
\end{figure}

Experimentally generating vortical states in a BEC requires adding angular momentum, phase winding, energy, or local flow, often starting from a stationary condensate formed through evaporative cooling. Not every such protocol produces turbulence; the point here is to identify the elementary routes by which vortices enter the fluid. Since the ground state of a BEC in the laboratory frame contains no vortices, vortices are created by driving, rotation, thermal quench, boundary injection, phase engineering, or direct excitation of compressible waves. Several methods for inducing vortices in quantum gases are shown in Fig.~\ref{fig:vmech}. The Thomas-Fermi approximation of the GPE gives useful intuition for mechanical driving: when $n(\mathbf{r})=[\mu - U(\mathbf{r})]/\varg$, time-modulating the potential strength, shape, or position directly controls the density landscape, local sound speed, and flow around obstacles.

Initial experiments sought to observe vortices directly, verifying the superfluid nature of the BEC and establishing the superfluid order parameter as central to condensate dynamics~\cite{matthews_vortices_1999,madison_vortex_2000}. The first observation of quantum vortices used phase and density engineering~\cite{matthews_vortices_1999}, while subsequent experiments shifted to mechanical forcing with magnetic or laser-induced potentials~\cite{madison_vortex_2000}. Reshaping and rotating the confining potential creates rapidly rotating condensates, both by rotating an initially condensed BEC~\cite{abo-shaeer_formation_2002,abo-shaeer_observation_2001} and by condensing from a rotating thermal cloud~\cite{haljan_driving_2001,engels_observation_2003,coddington_experimental_2004}. At fixed rotation frequency, or equivalently in the rotating frame with the corresponding angular-momentum constraint, these same-sign arrays minimize their energy by forming a vortex lattice~\cite{castin_bose-einstein_1999}; this chiral vortex matter is described in later sections and shown in Fig.~\ref{fig:chiralVortexMatter}.

Experimentally, excitation with a laser-induced potential can create vortex pairs as the superfluid locally exceeds the critical flow condition at the obstacle boundary, observed either through the onset of heating~\cite{raman_evidence_1999}, through direct vortex observation after time-of-flight~\cite{inouye_observation_2001,neely_formation_2010_review,kwon_observation_2016,samson_deterministic_2016,wilson_generation_2022,kwon_sound_2021}, or in situ~\cite{wilson_situ_2015}. More recently, expanded control from complex optical potentials has enabled the precise production, pinning, and placement of vortices in quasi-2D superfluids. The \textit{chopsticks} method, for example, uses the separation of two initially overlapped optical potentials to create pairs of vortices~\cite{samson_deterministic_2016}. Multiple pairs generated in this way were used to track vortex-dipole collisions~\cite{kwon_sound_2021}, and recently to create large-scale arrays of vortices in arbitrary configurations~\cite{gertjerenken_generating_2016,neely_melting_2024}.

The same control techniques support thermodynamic and topological routes to vortex formation. A rapid quench can freeze independent phase domains into defects~\cite{weiler_spontaneous_2008}, while in quasi-2D gases the Berezinskii-Kosterlitz-Thouless transition is associated with vortex-antivortex pair unbinding~\cite{hadzibabic_berezinskii-kosterlitz-thouless_2006}. Merging condensates with distinct phase, direct phase imprinting, topological pumping, and synthetic gauge fields impose phase windings that relax through vortices or persistent currents~\cite{scherer_vortex_2007,hernandez-rajkov_connecting_2024,del_pace_imprinting_2022,leanhardt_imprinting_2002,moon_thermal_2015,lin_synthetic_2009}.

The corresponding observables are no longer limited to density images. Vortex-resolved experiments measure positions and, in favorable geometries, circulation signs~\cite{seo_observation_2017,johnstone_evolution_2019,gauthier_giant_2019}; spectral experiments measure momentum distributions~\cite{henn_emergence_2009} or reconstruct incompressible/compressible energy spectra~\cite{johnstone_evolution_2019}; and tracer or reconstruction methods can access velocity statistics and structure functions~\cite{zhao_kolmogorov_2025}. Each observable has limitations. Imaging resolution can miss small vortex dipoles or core-scale sound; time-of-flight enlarges vortex cores but can obscure the in situ mapping between density and momentum; finite trap windows limit the available inertial range; vortex-sign detection for velocity field reconstruction is harder than density imaging. In weak-wave experiments, time-of-flight directly measures momentum-space density or occupation, from which wave-action or energy spectra are inferred using the dispersion relation and experimental calibration. Expansion does not provide comparably direct access to the incompressible kinetic-energy spectrum relevant for pure vortex turbulence. These limitations are an ongoing challenge, because flux-resolved turbulence requires the forcing, dissipation, spectra, and vortex content to be compared within the same controlled evolution.

Controlled few-vortex experiments~\cite{neely_observation_2010, middelkamp_guiding-center_2011, navarro_dynamics_2013, samson_deterministic_2016, gauthier_giant_2019, moon_thermal_2015, kwon_sound_2021, grani_mutual_2025_review, neely_melting_2024} provide a bottom-up route to the processes that become statistically organized in turbulence. Vortex lattices, bent vortex cores, real-time vortex precession, and observed vortex-line bending or reconnection identify the elementary dynamics--precession, bending, reconnection, and sound emission--that later appear inside turbulent ensembles~\cite{raman_vortex_2001,rosenbusch_dynamics_2002,freilich_real-time_2010,serafini_vortex_2017}.

\subsubsection{Vortex-sound coupling, reconnections, and damping}
Vortex processes transport and dissipate energy in ways unavailable to a smooth Euler fluid. In two dimensions, vortex dipoles scatter from other vortices and can annihilate upon sufficiently close approach, converting vortex kinetic energy into sound~\cite{capuzzi_annihilation_2008,prabhakar_annihilation_2013_review}. In three dimensions, vortex lines bend, reconnect, and form loops or vortex rings; reconnection changes the topology of a vortex tangle and launches strong local density waves~\cite{ogawa_study_2002}. The cusps generated by reconnection relax into Kelvin waves, providing a route from large-scale vortex motion to shorter-wavelength excitations and eventually phonon emission~\cite{fonda_direct_2014,kozik_vortex-phonon_2006,barenghi_anomalous_2006_review}.

Finite-temperature and openness add another layer to this Hamiltonian picture. Interaction between the condensate and any thermal non-condensate fraction damps vortex motion through exchange of energy and momentum. In helium this coupling is described phenomenologically as mutual friction, a force on the vortex proportional to its motion relative to the normal fluid~\cite{vinen_mutual_1957}. In dilute gases the microscopic origin is scattering between condensate excitations and the thermal cloud; phenomenologically it appears as dissipative GPE dynamics, stochastic projected-GPE descriptions, Zaremba-Nikuni-Griffin-type kinetic theories, vortex drift, and accelerated decay of vortex arrays or turbulent clusters~\cite{blakie_dynamics_2008,proukakis_finite-temperature_2008,fedichev_dissipative_1999-1,zhuravlev_dissipative_2001,kim_role_2016,stockdale_universal_2020,mehdi_mutual_2023}. Mutual friction is distinct from classical viscosity, but it plays an analogous organizing role by setting the timescale over which vortex energy is converted into heat, sound, and particle loss.

\begin{figure}[!tb]
\centering
	\includegraphics[width=\textwidth]{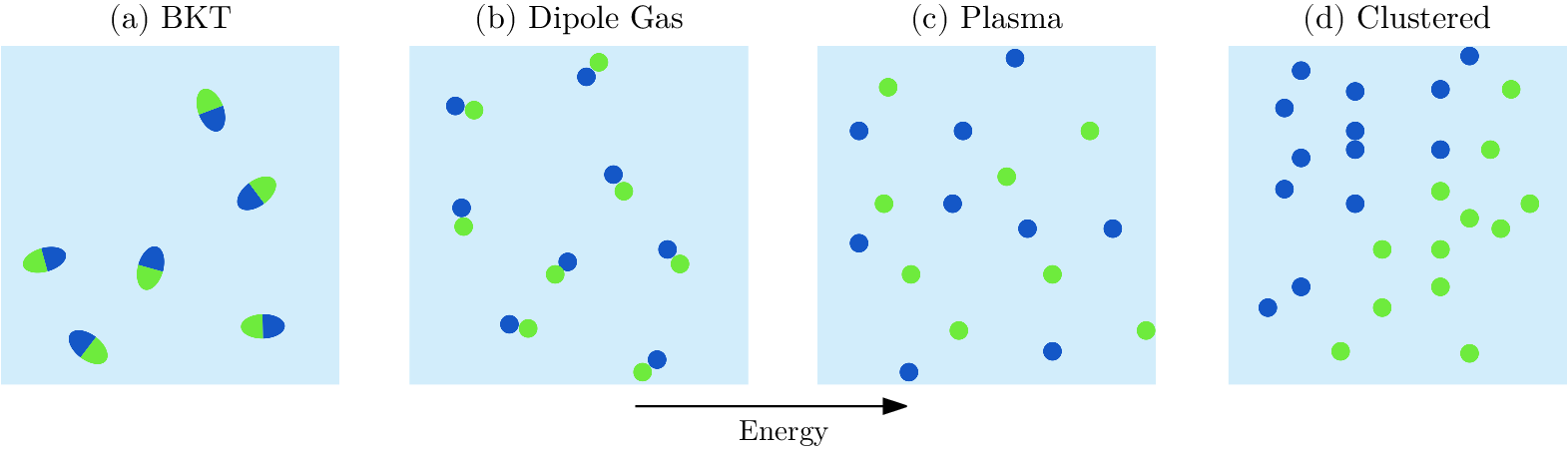}
	\caption{Equilibrium vortex phases in 2D, with vortex energy increasing left to right. (a) Phase-fluctuating BKT regime in strict 2D; In the effective 2D vortex regime: (b) low-energy vortex dipole gas; (c) uncorrelated vortex plasma; (d) high-energy clustered Onsager vortex phase}
    \label{fig:vortexGasPhases}
\end{figure}
\subsection{Equilibrium vortex phases in two dimensions}
Two-dimensional quantum vortex gases provide a useful equilibrium reference for interpreting vortex turbulence, because the point-vortex energy is controlled primarily by the signs and separations of the vortices. These phases are sketched in \fref{fig:vortexGasPhases}. In strict 2D confinement, the system enters a phase-fluctuating BKT regime. Weaker 2D confinement accesses a regime of effective 2D vortex motion with retained global phase coherence. Low energies dominated by vortex dipoles have little large-scale flow; increasing the energy results in an approximately uncorrelated vortex plasma with weak spatial order. At higher energies, Onsager vortex states emerge with large-scale circulation and negative absolute temperature in the point-vortex description~\cite{onsager_statistical_1949,billam_onsager-kraichnan_2014,simula_emergence_2014,yu_theory_2016,reeves_turbulent_2022}. These configurations have distinct signatures in the incompressible kinetic-energy spectrum, that we outline in Sec.~\ref{sec:spectra_fluxes}.

The BKT regime should be read as a separate thermal reference: there, an indefinite number of dipoles arise from finite-temperature phase fluctuations and unbind near the two-dimensional superfluid transition. The dipole gas, vortex plasma, and clustered states instead describe a neutral point-vortex gas of definite vortex number ordered by increasing conserved vortex energy. The same ordering is quantified in hard-wall Gross-Pitaevskii fields in \fref{fig:vortexGasSpectraCorrelations}, through the corresponding incompressible spectra and velocity correlations.

\begin{figure}[!t]
\centering
		\includegraphics[width=\textwidth]{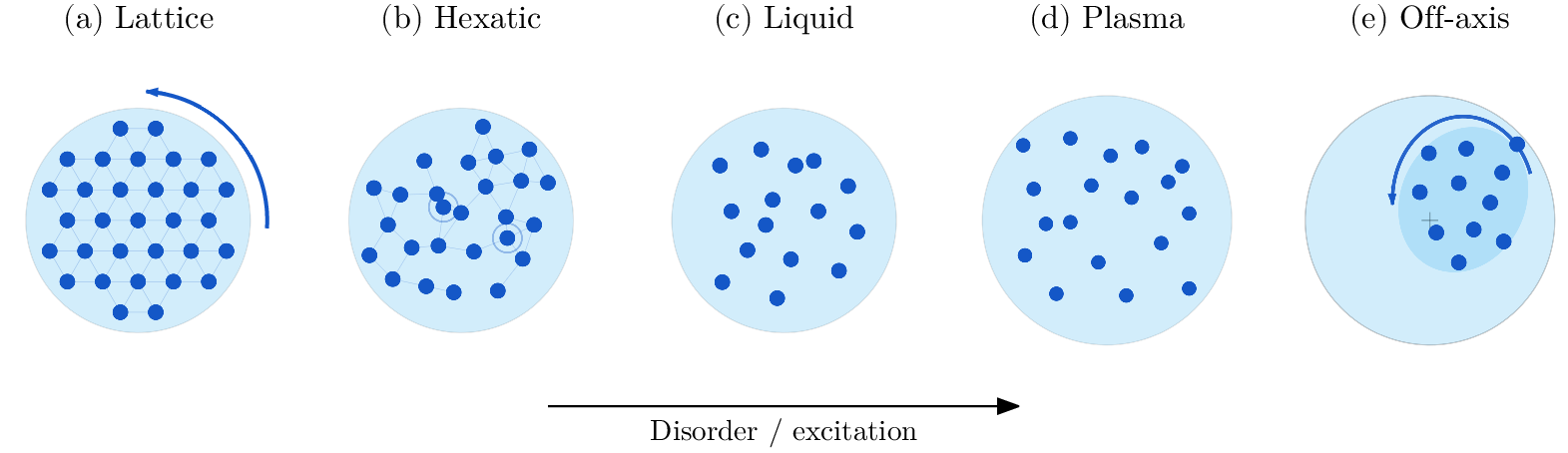}
		\caption{Schematic chiral vortex matter in 2D, with all vortices carrying positive counter-clockwise circulation. (a) At low excitation energy, same-sign vortices form an ordered triangular lattice with long-range positional and orientational order. (b) With increasing disorder the lattice can melt into a hexatic state, where local sixfold orientational order remains but positional order is disrupted. (c) Further excitation produces a liquid-like state with only short-range spatial correlations. (d) In the plasma-like regime, same-sign vortices are nearly uncorrelated and strongly disordered. (e) Confinement and finite angular momentum can also support off-axis same-sign vortex matter, which is distinct from the simple lattice-to-plasma melting sequence}
    \label{fig:chiralVortexMatter}
\end{figure}

This ordering is a useful baseline for distinguishing equilibrium vortex organization from driven nonequilibrium inverse transfer. A complementary organizing picture is provided by chiral vortex matter, shown in \fref{fig:chiralVortexMatter}. In this case the vortices have the same positive circulation, so the sequence is controlled by positional order, orientational order, and confinement rather than by the separation of vortex signs. The same-sign, non-neutral limit is closely related to guiding-centre plasma and point-vortex theories, where negative-temperature phase-transition behavior provides an early statistical-mechanical reference point~\cite{smith_phase-transition_1989}.

Spectrally, the lattice produces discrete features set by the intervortex spacing, the hexatic~\cite{sharma_thermal_2024} and liquid~\cite{sharma_thermal_2024,neely_melting_2024} regimes broaden these spatial-order peaks, and a disordered same-sign plasma gives a smoother spectrum whose low-$k$ enhancement depends on confinement and angular momentum.

\subsection{Spectra, spectral budgets, and fluxes}
\label{sec:spectra_fluxes}
The non-singular vortex core motivates a standard kinetic-energy decomposition, accounting for compressibility. The velocity field itself is singular at a vortex centre, but the density-weighted velocity field
\ali{
\mathbf{w}(\mathbf{r})&=\sqrt{n(\mathbf{r})}\mathbf{v}(\mathbf{r})
}
is free from pathology~\footnote{The density vanishes in the core as $n\sim r^2$, and hence $\mathbf{w}$ is regular.}. This is the field usually used to analyse the hydrodynamic kinetic energy in \eref{hydroH} for a compressible quantum fluid~\cite{nore_kolmogorov_1997}.

In Gross-Pitaevskii simulations, and in homogeneous or box-like experiments where Fourier shells are well defined, $\mathbf{w}$ can be separated by a Helmholtz decomposition~\footnote{The separation is nonlocal in real space, but local in Fourier space and may be implemented numerically via fast Fourier transforms.} into incompressible and compressible components,
\ali{
\mathbf{w}&=\mathbf{w}^{\rm i}+\mathbf{w}^{\rm c},&
\nabla\cdot\mathbf{w}^{\rm i}&=0,&
\nabla\times\mathbf{w}^{\rm c}&=0.
}
The shell-integrated incompressible ($i$ superscript) and compressible ($c$ superscript) kinetic-energy spectra are then
\ali{
E^{\rm i,c}(k)&=
\frac{m}{2}
\int_{|\mathbf{k}'|=k}
|\mathbf{w}^{\rm i,c}(\mathbf{k}')|^2\,dS_{\mathbf{k}'} .
}
The words ``incompressible'' and ``compressible'' in this notation refer to the transverse and longitudinal Helmholtz components of $\mathbf{w}$, not to the thermodynamic compressibility $\beta_s$ or to the strict hydrodynamic statement that a flow has no divergence: $\nabla\cdot\mathbf{v}=0$. The decomposition is a formal projection that separates solenoidal vortex kinetic energy from longitudinal sound-like kinetic energy inside a compressible quantum fluid~\cite{nore_kolmogorov_1997}. It should therefore not be read as saying that vortices and waves are dynamically independent, or that the underlying gas is physically incompressible. Its value is diagnostic: it asks how much kinetic energy resides in vortex degrees of freedom, how much resides in acoustic degrees of freedom, and how the two exchange energy during nucleation, annihilation, reconnection, and decay.

A spectrum specifies where energy or occupation resides in scale space; it does not by itself demonstrate that any quantity is being transported across scales. A cascade is a coherent nonequilibrium transport process. For a shell-integrated spectral density $X_a(k)$, where $a$ may denote an incompressible, compressible, wave-action, particle-number, or energy density in $k$-space, the time evolution can be written schematically as
\begin{equation}
    \partial_t X_a(k) = T_a(k) + F_a(k) - D_a(k) + C_a(k).
    \label{eq:spectral_budget}
\end{equation}
Here $T_a(k)$ is the nonlinear transfer among Fourier shells, $F_a(k)$ is injection, $D_a(k)$ is dissipation or loss, and $C_a(k)$ denotes conversion between sectors, for example vortex kinetic energy converted into sound during annihilation or reconnection~\cite{nore_kolmogorov_1997,bradley_energy_2012}. For a complete decomposition of a conserved quantity, conversion redistributes that quantity between sectors rather than creating it, so the integrated conversion terms cancel
\ali{
\sum_a\int_0^\infty C_a(k)\,dk&=0.
}
The precise separation of transfer and conversion is decomposition-dependent. With the convention
\begin{equation}
    \Pi_a(k) \equiv -\int_0^k T_a(q)\,dq ,
    \label{eq:cumulative_flux}
\end{equation}
positive $\Pi_a(k)$ denotes downscale transport and negative $\Pi_a(k)$ denotes upscale transport. An inertial-range cascade is indicated by an interval of $k$ between forcing and dissipation scales in which the direct injection and loss terms are negligible and $\Pi_a(k)$ is approximately constant. A power law is then supporting evidence for a cascade, not the definition of one. Clear phenomenology showing well-separated forcing and damping scales can provide strong supportive evidence for cascade physics, but it remains indirect unless accompanied by a flux measurement or an equivalent reconstruction of scale-to-scale transport.

This distinction is especially important in Gross-Pitaevskii turbulence because several mechanisms generate power laws without implying an inertial-range cascade. Single-vortex cores produce the ultraviolet $k^{-3}$ incompressible spectrum, finite vortex gases produce structure-factor-dependent spectra, equilibrium Onsager clustering enhances low-$k$ energy, and nonthermal coarsening can generate self-similar occupation spectra~\cite{bradley_energy_2012,kusumura_energy_2013,bradley_spectral_2022,nowak_nonthermal_2012,gazo_universal_2025}. In vortex-dominated regimes the relevant transport is usually expressed through the incompressible kinetic-energy budget $E^{\rm i}(k)$, while in weak-wave turbulence the natural transported quantities are wave action or particle number and energy, with cumulative fluxes denoted $\Pi_N(k)$ and $\Pi_E(k)$~\cite{zakharov_kolmogorov_1992,nazarenko_wave_2011,navon_synthetic_2019,zhu_direct_2023}. A cascade claim should therefore state the transported quantity, forcing scale, dissipation scale, flux direction, and the range over which the flux is approximately constant.

\begin{figure}[!htbp]
\centering
	\includegraphics[width=\textwidth]{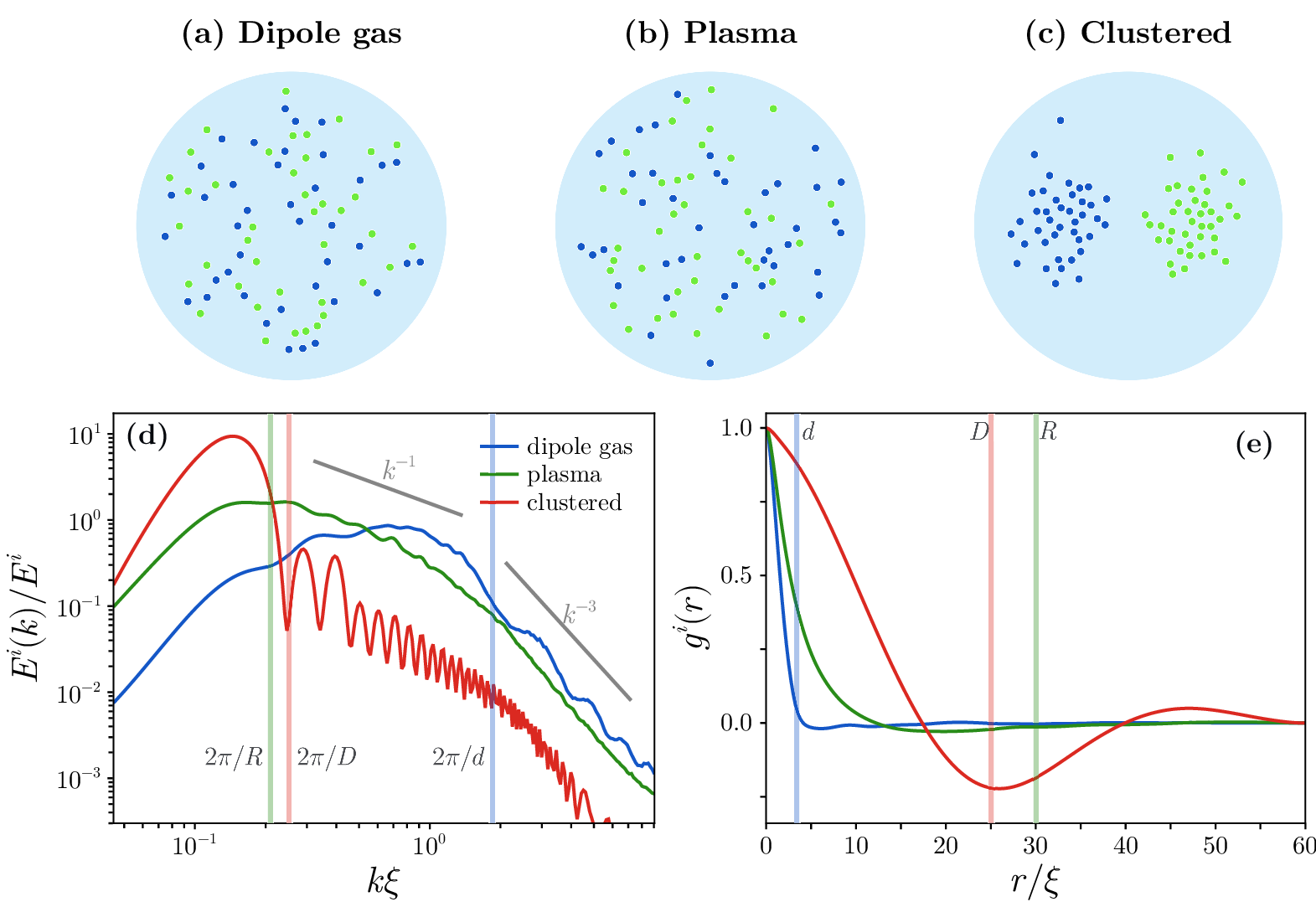}
	\caption{Gross-Pitaevskii vortex-gas diagnostics in a hard-wall disk trap. A Thomas-Fermi background density is imprinted with vortices (including images due to the boundary) to analyse spectra and correlations. (a) A low-energy vortex dipole gas consists of closely bound vortex-antivortex pairs, so far-field circulation is strongly cancelled. (b) An uncorrelated neutral plasma contains free positive and negative vortices with weak spatial order. (c) A clustered Onsager-like state is sampled from two independent elongated Gaussian clouds, one for each vortex sign, with their long axes tangential to the boundary; the resulting sign separation produces large-scale same-circulation flow. (d) The corresponding incompressible kinetic-energy spectra distinguish these regimes: dipoles suppress low-$k$ energy, the plasma approaches the incoherent single-vortex form over intermediate scales, and clustering enhances infrared spectral weight. Vertical guides mark the inverse length scales associated with the dipole size $d$, disk radius $R$, and cluster separation $D$. (e) The angle-averaged velocity two-point correlations provide the real-space counterpart of the spectra, showing rapid dipole decorrelation, weak plasma correlations, and long-range correlations from same-sign clustering}
    \label{fig:vortexGasSpectraCorrelations}
\end{figure}

For a two-dimensional distribution of $N$ well separated point vortices the incompressible spectrum has a simple form,
\ali{
E^{\rm i}(k)&=
E_1(k)
\left[
N+2\sum_{i<j} q_i q_j J_0(k r_{ij})
\right],
}
where $q_i=\pm 1$, $r_{ij}=|\mathbf{r}_i-\mathbf{r}_j|$, and $E_1(k)$ is the single-vortex spectrum~\cite{bradley_energy_2012}. With the healing-length convention used here,
\ali{
E_1(k)&\propto k^{-1}, \qquad k\xi\ll 1,\\
E_1(k)&\propto k^{-3}, \qquad k\xi\gg 1.
}
The factor in square brackets is therefore the vortex-gas structure factor. The same spectral information can also be represented in real space: for an isotropic two-dimensional flow, the angle-averaged velocity two-point correlation function and the kinetic-energy spectrum are related by a Bessel/Hankel transform~\cite{bradley_spectral_2022}. The spectra and velocity-correlation panels in \fref{fig:vortexGasSpectraCorrelations} should therefore be read as complementary diagnostics of the same vortex correlations. A tightly bound vortex-antivortex dipole has suppressed infrared incompressible energy because the far-field circulation cancels; its spectrum rolls over below $k\sim d^{-1}$, where $d$ is the dipole size, and approaches the incoherent single-vortex sum above that scale. A neutral random vortex plasma has weak pair correlations and approximately gives $E^{\rm i}(k)\sim N E_1(k)\propto Nk^{-1}$ for $R^{-1}\ll k\ll \xi^{-1}$, with $R$ the system size~\cite{kusumura_energy_2013}; exact neutrality suppresses the strict $k\to0$ limit in a finite system. An Onsager-clustered state enhances the low-$k$ spectrum because same-sign correlations add constructively through the $q_i q_j J_0(k r_{ij})$ term. A $k^{-5/3}$ range is therefore not the generic spectrum of every negative-temperature equilibrium state. It is better regarded as a correlated nonequilibrium, or inverse-cascade-associated, signature when it is produced dynamically by driven or decaying turbulent evolution~\cite{novikov_dynamics_1975,bradley_energy_2012,skaugen_origin_2017}.

These budget variables also sharpen how wave-turbulent regimes are named: their cascade content is carried by occupation spectra, phase correlations, and measured or reconstructed fluxes of wave action or particles and energy, denoted for example by $\Pi_N(k)$ and $\Pi_E(k)$, rather than by spectral slope alone~\cite{zakharov_kolmogorov_1992,nazarenko_wave_2011,navon_emergence_2016,zhu_direct_2023}. The most useful regime labels therefore combine dimensionality and spectral exponents with forcing scale, forcing amplitude relative to $\mu$, dissipation scale, compressibility, flux direction, and which spectral flux is actually measured or inferred. Table~\ref{tab:power-law-status} summarizes the main power-law predictions used in this chapter and the current measurement status in atomic gases.

\begin{table}[t]
\centering
\caption{Key power-law predictions for atomic-gas turbulence and their measurement status. The table separates cascade evidence from baseline vortex-core or coarsening scalings, because the same exponent can have different physical meanings in different regimes. The listed exponents are therefore diagnostic labels; cascade identification requires corresponding evidence for scale-to-scale flux or an equivalent transport measurement.}
\label{tab:power-law-status}
\scriptsize
\setlength{\tabcolsep}{3pt}
\renewcommand{\arraystretch}{1.18}
\setlength{\extrarowheight}{1pt}
\begin{tabularx}{\textwidth}{@{}>{\raggedright\arraybackslash}p{0.22\textwidth}>{\raggedright\arraybackslash}p{0.19\textwidth}>{\raggedright\arraybackslash}p{0.19\textwidth}>{\raggedright\arraybackslash}X@{}}
\hline
Theory / regime & Power-law prediction & What it diagnoses & Measurement status \\
\hline
Single-vortex spectrum~\cite{bradley_energy_2012,reeves_inverse_2013} & $E_1(k)\propto k^{-1}$ for $k\xi\ll 1$; $E_1(k)\propto k^{-3}$ for $k\xi\gg 1$ & Baseline quantized-vortex structure, not a cascade & Used as the reference spectrum in GPE and vortex-resolved analyses; observing the ultraviolet $k^{-3}$ tail alone does not establish turbulent transport \\
3D quasi-classical vortex turbulence~\cite{kolmogorov_local_1941,nore_kolmogorov_1997,kobayashi_kolmogorov_2005,barenghi_types_2023} & $E^{\rm i}(k)\propto \varepsilon^{2/3}k^{-5/3}$ over scales larger than the intervortex spacing & Evidence consistent with a direct incompressible-energy cascade; vortex polarization or bundling requires independent real-space diagnostics & Strong support from helium and GPE simulations; atomic-gas evidence is suggestive, but a fully vortex-line- and flux-resolved measurement remains open \\
2D inverse vortex cascade~\cite{bradley_energy_2012,reeves_inverse_2013,johnstone_evolution_2019} & $E^{\rm i}(k)\propto k^{-5/3}$ below the forcing scale during upscale transfer & Evidence consistent with inverse incompressible-energy transport; same-sign clustering requires vortex-resolved correlation data & GPE simulations predict the cascade; experiments have observed Onsager clustering and inferred intermediate-time $k^{-5/3}$-like spectra, but observing vortex-energy flux remains open \\
Classical direct wave cascade~\cite{zakharov_kolmogorov_1992,nazarenko_wave_2011,navon_emergence_2016,navon_synthetic_2019,galka_emergence_2022,zhu_direct_2023} & Kolmogorov--Zakharov occupation spectra $n(k)\propto k^{-\gamma}$, with $\gamma\simeq 3.5$ for the particle-like four-wave direct cascade & Downscale transport of wave energy through particle-like Gross--Pitaevskii modes & Momentum-space power laws have been measured in 2D and 3D box gases, with reported $\gamma$ values near $2.9$--$3.5$; engineered dissipation gives flux evidence, while exact slopes remain geometry- and definition-dependent \\
Classical weak-wave inverse particle cascade~\cite{zhu_direct_2023,karailiev_observation_2024} & In 2D, the formal weak-wave prediction is $n(k)\propto k^{-4/3}$ for an inverse particle or wave-action flux & Upscale transport of occupation from the forcing scale & A driven homogeneous 2D gas measured $\gamma=1.55(15)$, close to the formal exponent, although the weak-wave solution has a self-consistency limitation in 2D; direct low-$k$ flux measurement remains open \\
Nonthermal coarsening / fixed-point scaling~\cite{nowak_nonthermal_2012,gazo_universal_2025,karailiev_observation_2024} & Low-$k$ $n(k)\sim [1+(k/k_0)^\kappa]^{-1}$ with $\kappa\simeq 3$ & Strongly interacting large-scale coarsening rather than weak-wave cascade transport & Observed in isolated 2D coarsening and in the low-$k$ sector of driven inverse-wave experiments; interpretation as a stabilized nonthermal fixed point remains cautious \\
Velocity structure functions~\cite{kolmogorov_local_1941,zhao_kolmogorov_2025} & $S_2(r)\sim r^{2/3}$ and $S_3(r)\sim r$, with intermittency in higher moments & Real-space Kolmogorov statistics of velocity increments & Recently measured with spinor-impurity tracers in turbulent 2D BECs, complementing spectra and vortex-position diagnostics \\
Vinen/ultraquantum decay~\cite{vinen_introduction_2006,barenghi_types_2023,morris_observation_2026_review} & Random-tangle vortex-line density $L(t)\propto t^{-1}$; quasi-classical decay often gives $L(t)\propto t^{-3/2}$ & Distinguishes random vortex tangles from large-scale polarized flow & Helium provides the benchmark; atomic-gas measurements are emerging, including a 2026 preprint reporting Vinen-like decay in a uniform 3D BEC \\
\hline
\end{tabularx}
\end{table}

\subsection{Wave turbulence and weak-wave dynamics}
The weak-wave regime of the turbulent GPE consists of small-amplitude density and phase waves on a nearly uniform condensate, rather than quantized vortices~\cite{zhu_direct_2023,zakharov_kolmogorov_1992}. The Bogoliubov dispersion is a quantum-superfluid property, but the turbulent redistribution of highly occupied modes is described by a classical wave-kinetic equation in the short-wavelength regime corresponding to nearly free particle evolution. Energy is transported by weakly nonlinear fluctuations of the condensate background, and the central questions concern resonant mode coupling, kinetic spectra, wave-action and energy fluxes, and the crossover to vortex nucleation when the wave field becomes sufficiently activated. We therefore refer to this regime as classical wave turbulence in a quantum gas, reserving quantum turbulence for regimes with active topological excitations.

The relevant small parameter is the nonlinear frequency shift compared with the linear wave frequency. When density and phase fluctuations are weak, modes remain close to the Bogoliubov dispersion while resonant mode coupling slowly redistributes occupation through wavenumber space. This separates the fast oscillation time from the slower cascade time and permits a kinetic description of wave action and energy transfer~\cite{zakharov_kolmogorov_1992,nazarenko_wave_2011}. Momentum distributions, occupation spectra, and phase correlations are therefore natural diagnostics of weak-wave turbulence, but the cascade interpretation still requires the flux or equivalent scale-to-scale-transport information described in Sec.~\ref{sec:spectra_fluxes}.

For dilute-gas BECs this regime is most cleanly accessed in homogeneous or box-like traps, where the sound speed and density are nearly uniform and geometric inhomogeneity does not dominate the spectrum. At long wavelength the excitations are acoustic phonons, while near $k\xi\sim 1$ the dispersion crosses over toward particle-like behavior. The direct energy cascade observed by Navon \textit{et al.} occupies this particle-like range and is a Kolmogorov--Zakharov solution of the four-wave kinetic equation~\cite{navon_emergence_2016,navon_synthetic_2019} for isotropic interactions (with a logarithmic correction~\cite{zhu_direct_2023}). When the wave amplitude becomes comparable to the chemical-potential scale, density depletions and phase gradients can nucleate vortices, driving the system into a mixed classical-wave and quantum-vortex regime~\cite{navon_emergence_2016,navon_quantum_2021,fischer_regimes_2025}.

\section{Experimental turbulent evolution and transport}
Experiments make the regime map operational by preparing nonequilibrium many-vortex, wave-dominated, or mixed states and measuring their spectra, correlations, fluxes, and decay. The examples below therefore emphasize turbulent evolution and transport, rather than the isolated vortex-creation mechanisms reviewed above.

Dimensionality remains the first organizing variable, but the dynamical protocol is equally important. Decaying turbulence starts from a prepared nonequilibrium state and relaxes without sustained injection; spectra, vortex number, vortex-line density, correlations, and energy then evolve toward thermal equilibrium. Forced-dissipative turbulence instead maintains injection and removal of energy or particles at controlled scales, making the central question whether a statistically steady flux connects the forcing and dissipation ranges.

\subsection{Two-dimensional vortex turbulence and Onsager clustering}

Effectively two-dimensional gases give the cleanest connection between experiments and point-vortex theory, because vortex lines cannot bend and vortex positions can be compared directly with the model degrees of freedom. Optical potentials have enabled controlled vortex injection through superfluid breakdown around barriers, including vortex-dipole formation and small clusters in oblate condensates~\cite{neely_observation_2010,samson_deterministic_2016,kwon_periodic_2015,kwon_observation_2016}. Programmable variants of the same control now allow deterministic vortex placement and collision studies~\cite{kwon_sound_2021}. These protocols define the injection mechanisms; the corresponding many-vortex states are often best described as decaying relaxation or turbulent equilibration, rather than continuously forced steady turbulence. Neely \textit{et al.} subsequently observed large-scale coherent structures forming in decaying 2D quantum turbulence~\cite{neely_characteristics_2013}. Supporting Gross-Pitaevskii simulations showed energy moving upscale and a Kolmogorov-like incompressible spectrum, but not a direct measurement of a constant inertial-range flux.

\begin{figure}[!htbp]
\centering
\includegraphics[width=\textwidth]{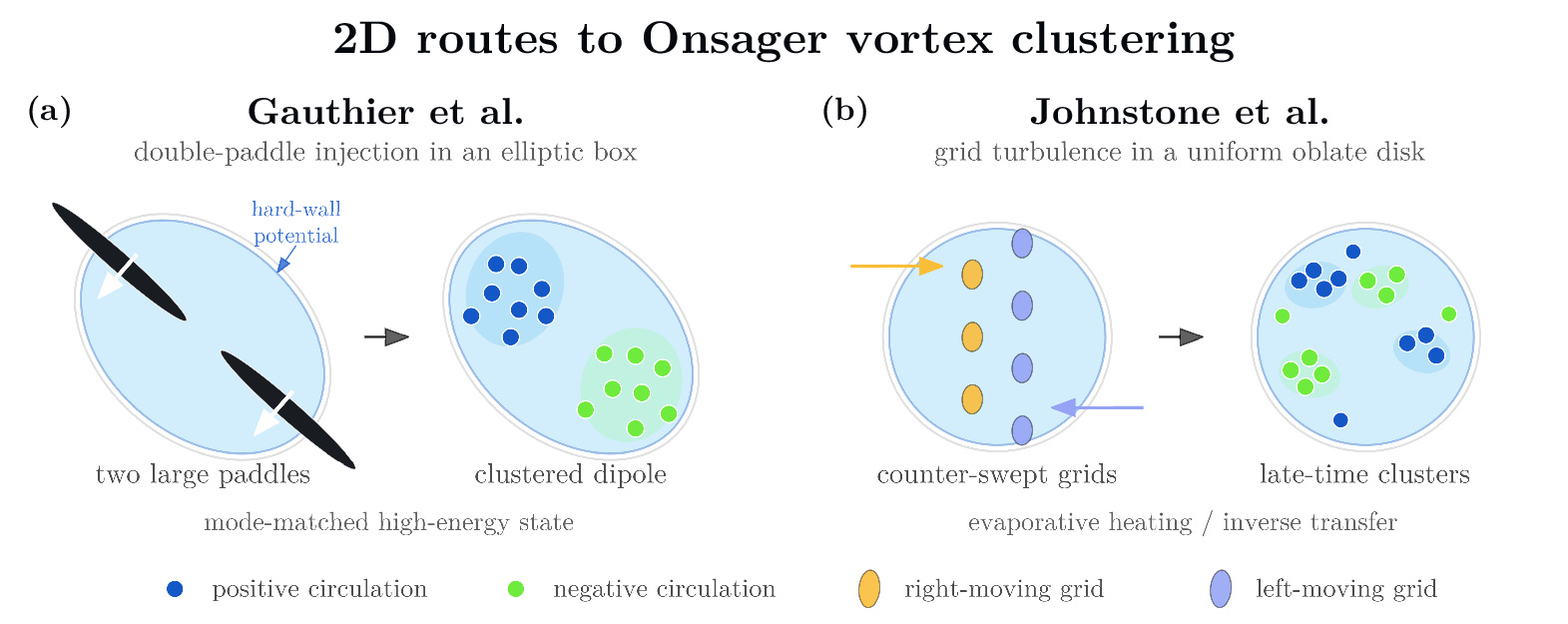}
	\caption{Schematic protocols for producing Onsager clustering in 2D quantum gases. (a) In the Gauthier \textit{et al.} experiment~\cite{gauthier_giant_2019}, two extended optical paddles were swept through an elliptical box trap, mode matching the high-energy vortex dipole and directly producing separated same-circulation clusters. (b) In the Johnstone \textit{et al.} experiment~\cite{johnstone_evolution_2019}, interleaved optical grids swept through a uniform oblate disk from opposite directions; vortex-antivortex annihilation and inverse transfer increase the energy per remaining vortex, giving a clustered negative-temperature vortex gas. Blue and green points denote opposite vortex circulations; black and colored ellipses denote repulsive optical barriers, with gold and violet marking the two grid directions}
	    \label{fig:clusterStirringProtocols}
\end{figure}

Gauthier \textit{et al.} and Johnstone \textit{et al.} observed Onsager clustering and long-lived large-scale circulation~\cite{gauthier_giant_2019,johnstone_evolution_2019}. In the latter case an intermediate-time $k^{-5/3}$ range was inferred from vortex-resolved spectra~\cite{johnstone_evolution_2019}, consistent with the low-$k$ enhancement expected when same-sign correlations add constructively. The two routes are sketched in \fref{fig:clusterStirringProtocols}: Gauthier \textit{et al.} injected a high-energy clustered dipole with large paddles in an elliptic box, whereas Johnstone \textit{et al.} used swept optical grids to prepare turbulent vortex gases whose vortex populations and energy evolve toward clustered, negative-temperature states. These studies were enabled by methods for determining vortex sign as well as position through Bragg scattering~\cite{seo_observation_2017}. Their strongest evidence is therefore vortex-resolved organization, not a direct measurement of vortex-energy flux. A chiral system of same-sign vortices provides a cleaner relaxation problem, with different initial conditions evolving toward equilibrium states predicted by the point-vortex model~\cite{reeves_turbulent_2022}.

Fluid instabilities provide another route to turbulent initial conditions. Kelvin-Helmholtz roll-up, for example, can seed vortex-rich states that subsequently decay~\cite{baggaley_kelvin-helmholtz_2018}. Recent annular and shear-flow experiments have realized this sequence, resolving both the instability and the ensuing decaying turbulence with vortex clustering~\cite{hernandez-rajkov_connecting_2024,simjanovski_shear-induced_2025}.

\subsection{Vortex and wave turbulence in three-dimensional box traps}
Optical box traps~\cite{gaunt_bose-einstein_2013,chomaz_emergence_2015,gauthier_direct_2016,navon_quantum_2021} sharpen the quantitative meaning of three-dimensional vortex-line and wave-turbulence measurements. A nearly uniform density gives well-defined values of $c$, $\xi$, forcing wavenumber $k_f$, and spectral shells, so measured momentum and occupation spectra can be compared more directly with homogeneous Gross-Pitaevskii or wave-kinetic theory~\cite{navon_emergence_2016,navon_quantum_2021}. Simulations complement this by separating compressible and incompressible kinetic energy, allowing the wave component to be tracked before, during, and after vortex nucleation.

\begin{figure}[!htbp]
\centering
\includegraphics[width=\textwidth]{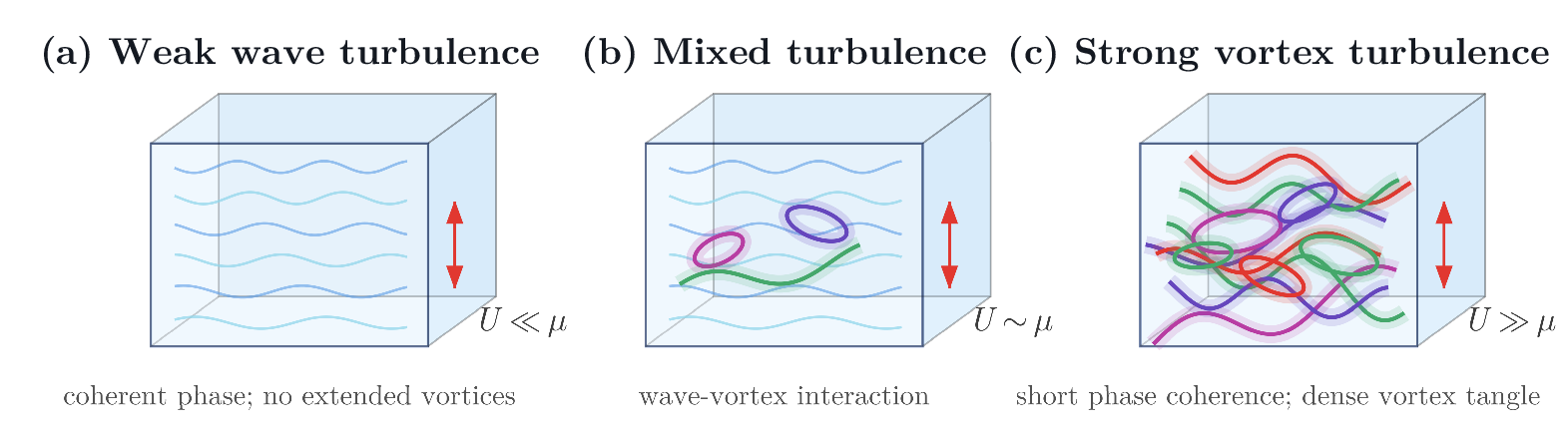}
		\caption{Schematic of forcing regimes for a box-confined three-dimensional Bose gas under shaking. Increasing the shaking amplitude $U$ relative to the chemical potential $\mu$ moves the fluid from (a) classical-field wave turbulence ($U\ll\mu$), where particle-like Bogoliubov modes dominate and extended vortices are absent, through (b) a mixed classical-wave and quantum-vortex crossover ($U\sim\mu$), to (c) strong vortex turbulence ($U\gg\mu$), a topological quantum-turbulence regime where dense vortex lines and short-range phase coherence dominate the bulk. The diagram connects the box-shaking cascade experiment of Navon \textit{et al.}~\cite{navon_emergence_2016} with steady-state Gross--Pitaevskii simulations~\cite{fischer_regimes_2025}}
    \label{fig:forced3DRegimes}
\end{figure}

In the pioneering box experiment~\cite{navon_emergence_2016}, energy was injected at large scales by applying a periodically driven linear potential along one axis. The direct cascade develops over particle-like wavenumbers and is described by classical four-wave kinetics, while the quantum gas supplies the coherent matter-wave field and its tunable dispersion. The resulting box-shaking geometry also provides a regime diagram: the forcing amplitude $U$ can be compared with the characteristic superfluid energy $\mu$, while spectra, phase correlations, and vortex content distinguish classical-field wave turbulence from a mixed regime and, under sufficiently strong forcing, strong vortex turbulence. This progression is sketched in \fref{fig:forced3DRegimes}.

Continuous excitation can be supplied by moving a wall~\cite{navon_emergence_2016,navon_synthetic_2019} or by imposing a controlled external potential~\cite{ville_sound_2018}. The trap depth or engineered loss can also set a high-$k$ dissipation scale $k_\textrm{cutoff}$~\cite{navon_synthetic_2019}. This makes it possible to compare injected power, particle loss, and the fluxes $\Pi_N(k)$ and $\Pi_E(k)$ through selected shells, rather than inferring cascades only from the slope of a spectrum. Box gases have therefore been central to measurements of direct and inverse transport in 3D~\cite{navon_synthetic_2019,dogra_universal_2023,martirosyan_equation_2026} and 2D~\cite{galka_emergence_2022,karailiev_observation_2024}.

Box-trap control has also enabled tests of the Gross-Pitaevskii description far from equilibrium, including measurements of a turbulent equation of state~\cite{dogra_universal_2023,martirosyan_equation_2026}. In tunable species such as $^{39}$K, interaction strength can be varied while keeping the geometry fixed; this has been used to observe subdiffusive scaling for weak interactions and the emergence of developed wave turbulence at stronger interactions in the presence of weak disorder~\cite{martirosyan_observation_2024}. Strongly driven harmonically trapped BECs gave early evidence for cascade-like energy redistribution in 3D~\cite{henn_emergence_2009}, while a recent preprint reports Vinen/ultraquantum turbulence in a uniform 3D BEC with vortex-line observables closer to the helium tradition~\cite{morris_observation_2026_review}. The remaining experimental challenge is to connect vortex-line density decay, reconnections, Kelvin-wave cascades, sound emission, and momentum-space fluxes in one resolved measurement sequence.

These examples show why atomic-gas turbulence cannot be classified by spectral slopes alone. Vortex clustering, controlled vortex-sound conversion, flux reconstruction, velocity statistics, and decay laws give complementary evidence for turbulent transport. Table~\ref{tab:becturbulence-experiments} summarizes representative BEC quantum-turbulence experiments by their dominant phenomenology and diagnostic evidence.

\begin{table}[!htbp]
\centering
\caption{Representative atomic-gas superfluid-turbulence experiments classified by phenomenology and diagnostic evidence. The table emphasizes that turbulence claims are not established by power laws alone: forcing and damping scales, vortex clustering, fluxes, velocity statistics, decay laws, and vortex-sound conversion provide complementary evidence.}
\label{tab:becturbulence-experiments}
\scriptsize
\setlength{\tabcolsep}{3pt}
\renewcommand{\arraystretch}{1.22}
\setlength{\extrarowheight}{1pt}
\begin{tabularx}{\textwidth}{@{}>{\raggedright\arraybackslash}p{0.19\textwidth}>{\raggedright\arraybackslash}p{0.22\textwidth}>{\raggedright\arraybackslash}p{0.21\textwidth}>{\raggedright\arraybackslash}X@{}}
\hline
Experiment / class & Geometry and forcing & Key phenomenology & Diagnostic evidence \\
\hline
Oscillating trapped BEC~\cite{henn_emergence_2009} & 3D harmonic trap; strong oscillatory trap excitation & Disordered vortex tangle; early atomic-gas turbulence evidence & Expansion images and vortex disorder; cascade-like redistribution inferred, without direct inertial-range flux \\
Decaying 2D vortex turbulence~\cite{neely_characteristics_2013} & Highly oblate BEC; obstacle-generated vortex dipoles and clusters & Decaying vortex turbulence; large-scale coherent structures & Experimental vortex imaging; GPE simulations support upscale incompressible-energy transfer and a $k^{-5/3}$-like spectrum, not a direct flux measurement \\
Giant vortex clusters~\cite{gauthier_giant_2019} & Uniform 2D box; double-paddle injection in an elliptic trap & Onsager clustering; large-scale same-sign circulation & Vortex positions, circulation signs, and cluster formation; evidence through negative-temperature vortex organization \\
Large-scale flow from turbulence~\cite{johnstone_evolution_2019} & Uniform oblate 2D BEC; swept optical grids & Decaying vortex gas; inverse transfer; Onsager clustering & Vortex-resolved spectra, signs, vortex number, and energy evolution; intermediate-time $k^{-5/3}$-like spectrum \\
Programmable vortex collider~\cite{kwon_sound_2021} & Quasi-2D programmable BEC; deterministic vortex placement and collision & Vortex--antivortex annihilation; vortex energy converted to sound & Vortex trajectories, annihilation events, and emitted sound; process-level evidence rather than a cascade exponent \\
Annular shear and Kelvin--Helmholtz flows~\cite{hernandez-rajkov_connecting_2024,simjanovski_shear-induced_2025} & Annular or quasi-2D superfluid; shear flow or vortex-array instability & Kelvin--Helmholtz roll-up; decaying turbulence and clustering & Instability growth, roll-up, vortex clustering, and decay; power-law evidence secondary \\
Emergent classical-field cascade~\cite{navon_emergence_2016} & 3D uniform box gas; shaking or moving-boundary excitation plus high-$k$ dissipation & Particle-like four-wave cascade, with  crossovers to mixed and strong vortex turbulence with increasing drive & Momentum distributions, occupations, and phase coherence; self-similar spectral evolution \\
Synthetic dissipation and fluxes~\cite{navon_synthetic_2019} & 3D box gas; controlled drive plus high-$k$ dissipation & Classical-field direct energy transport with controlled dissipation & Injected power, atom loss, spectra, and reconstructed $\Pi_N(k)$ and $\Pi_E(k)$ \\
Homogeneous 2D wave turbulence~\cite{galka_emergence_2022} & Homogeneous 2D Bose gas; driven weak-wave excitation & Isotropization, dynamic scaling, and 2D wave turbulence & Momentum distributions, occupation spectra, and time-dependent scaling collapse \\
Inverse turbulent-wave cascade~\cite{karailiev_observation_2024} & Driven homogeneous 2D Bose gas; finite-$k$ forcing & Inverse wave transport toward larger length scales; coarsening & Occupation transfer, long-time momentum distributions, and scaling/coarsening evidence \\
Turbulent equation of state~\cite{dogra_universal_2023,martirosyan_equation_2026} & Uniform quantum gas in a wave-turbulent regime & Turbulent thermodynamics; universal equation of state & Energy, particle number, momentum spectra, and thermodynamic collapse \\
Velocity structure functions~\cite{zhao_kolmogorov_2025} & Turbulent 2D BEC with spinor-impurity velocity probes & Kolmogorov-like velocity statistics; intermittency signatures & Velocity increments and structure functions; evidence complementary to spectra and vortex coordinates \\
\hline
\end{tabularx}
\end{table}

The classification in Table~\ref{tab:becturbulence-experiments} returns to the diagnostic distinction used throughout this chapter. Power-law spectra are useful, but they are only one form of evidence. In finite quantum gases, vortex-resolved correlations, sign clustering, measured fluxes, decay laws, structure functions, and controlled vortex-sound conversion can be equally important for identifying the turbulent regime.

\section{Recent advances and emerging directions}

Recent work has improved diagnostic measurements for atomic-gas turbulence. Disordered vortices and fitted power laws remain useful signatures, and focus has increasingly moved to understanding forcing, loss, occupation spectra, vortex-resolved observables, velocity statistics, and flux-sensitive diagnostics within controlled Gross--Pitaevskii fluids~\cite{amette_estrada_turbulence_2022,zhao_kolmogorov_2025,zhu_direct_2023,barenghi_is_2008,barenghi_types_2023}. This sharper standard also separates classical-field cascade phenomenology from topological quantum turbulence: smooth waves carry classical wave-kinetic transport on a quantum-superfluid dispersion, whereas vortices carry quantized circulation and can exchange energy with the wave field.

Classical-field wave-turbulent transport provides a clear watershed in the development of atomic-gas turbulence studies. High-resolution Gross--Pitaevskii and wave-kinetic calculations predict simultaneous direct and inverse cascades in driven Bose gases~\cite{zhu_direct_2023}. Experiments in homogeneous box gases have followed the build-up of wave turbulence, introduced controlled high-$k$ dissipation, reconstructed particle and energy fluxes, and measured turbulent equations of state~\cite{navon_emergence_2016,navon_synthetic_2019,dogra_universal_2023}. The Navon \emph{et al.} direct cascade experiment lies in the particle-like dispersion range and realizes a Kolmogorov--Zakharov cascade of the four-wave kinetic equation; it is therefore classical in mechanism even though it occurs in a quantum gas. In two dimensions, driven homogeneous gases show isotropization, dynamic scaling, and inverse turbulent-wave transport from the forcing scale toward larger length scales, with long-time spectra that connect weak-wave turbulence to nonthermal coarsening phenomenology~\cite{galka_emergence_2022,karailiev_observation_2024,gazo_universal_2025}. These measurements make the transported quantity and the dissipation channel part of the experimental evidence. This is in contrast to vortex turbulence, where often the cascade interpretation has relied on a spectral exponent alone.

Vortex turbulence has advanced through complementary diagnostics. Controlled vortex colliders show vortex energy converted into sound~\cite{kwon_sound_2021}; annular and shear-flow geometries connect fluid instabilities with vortex-rich decaying turbulence~\cite{hernandez-rajkov_connecting_2024}; and vortex-sign-resolved experiments connect Onsager clustering to the low-$k$ enhancement of the incompressible spectrum~\cite{johnstone_evolution_2019}. These measurements do not all establish cascades in the same sense. Their common role is to relate elementary vortex processes, vortex correlations, and spectral signatures within the finite, compressible geometry of an atomic gas.

Velocity-field statistics provide a further bridge to classical turbulence. Spinor-impurity tracer measurements have enabled structure functions in turbulent two-dimensional Bose gases, giving Kolmogorov-like scaling and intermittency signatures in a form directly comparable with classical turbulence~\cite{zhao_kolmogorov_2025}. Three-dimensional vortex-line observables remain a central piece of the turbulence program. Recent work is beginning to approach the helium-style language of vortex-line density and Vinen/ultraquantum decay in uniform atomic gases; a preprint reports randomly oriented vortex-line imprints and decay consistent with Vinen turbulence~\cite{morris_observation_2026_review}. 

Going beyond scalar weakly interacting condensates, the contact-interaction Gross-Pitaevskii fluid forms the reference case against which new hydrodynamic couplings, internal degrees of freedom, and dissipation mechanisms can be compared.

\subsection{Beyond contact interactions: anisotropic and driven quantum fluids}

The scalar contact-interaction degenerate Bose gas is the clean baseline for quantum-gas turbulence. In this setting, quantized circulation, compressibility, vortices, sound, and weak-wave turbulence can be separated most transparently, and the diagnostic language of $E^{\rm i}(k)$, $E^{\rm c}(k)$, vortex structure factors, occupation spectra, and fluxes has a direct meaning. The broader question is which of these concepts remain universal when the equation of state, interaction kernel, internal order parameter, or conservation laws are changed.

For nonlocal two-body interactions the interaction energy is no longer locally $\varg n^2/2$. It has the form
\ali{\label{nonlocal_int}
E_{\rm int}&=\frac{1}{2}\int d^D\mathbf{r}\,d^D\mathbf{r}'\,n(\mathbf{r})V(\mathbf{r}-\mathbf{r}')n(\mathbf{r}') .
}
The Bogoliubov spectrum then depends on the Fourier transform $\tilde V(\mathbf{k})$, not only on a scalar coupling constant. In dipolar gases this can make the sound speed and roton softening direction-dependent and can modify resonant wave couplings and weak-wave dynamics~\cite{chomaz_dipolar_2022,ticknor_anisotropic_2011}. Circularly averaged spectra can therefore hide key physics of nonlocal and anisotropic fluids.

Dipolar gases also expose a useful tension between two kinds of long-range physics. In a two-dimensional contact superfluid, the long-range vortex-vortex interaction comes from phase-gradient kinetic energy. A dipolar or otherwise nonlocal interaction does not replace this hydrodynamic/topological contribution: circulation still controls the sign-dependent vortex flow. But a vortex is also a finite-core density depletion. Because the density field participates in the nonlocal interaction energy \eref{nonlocal_int}, a vortex carries a nonlocal density defect in addition to its phase winding. Thus the hydrodynamic part is topological, sign-dependent, and circulation-mediated, while the dipolar correction is density-mediated, core-sensitive, anisotropic, and dependent on polarization direction and compressibility~\cite{bland_vortices_2023,casotti_observation_2024}.

These established nonlocal and anisotropic effects motivate, but do not yet by themselves establish, a general phenomenology of dipolar vortex turbulence. A useful program is therefore to test angular diagnostics such as $E(k,\theta)$, $n(k,\theta)$, anisotropic structure factors, and direction-dependent fluxes against their circularly averaged counterparts $E(k)$ and $n(k)$, while treating interaction-dependent changes to vortex trajectories, annihilation, clustering, and lattice geometry as questions requiring direct system-specific evidence.

Spinor and multicomponent fluids add internal turbulent degrees of freedom rather than only modifying the scalar equation of state. Mass flow can coexist with spin flow, countersuperflow, domain walls, half-quantum vortices, skyrmions, and spin-wave turbulence~\cite{stamper-kurn_spinor_2013,kawaguchi_spinor_2012}. In these systems we can ask whether cascade phenomenology should be attached to a single velocity field or to coupled transport among mass, spin~\cite{fujimoto_spin_2013-1,lee_energy_2026_review} and relative-phase sectors. Fermionic superfluids introduce a different set of constraints: BEC-BCS crossover physics, pair breaking, vortex-core quasiparticles, and critical-velocity mechanisms that need not reduce to Gross-Pitaevskii vortex shedding~\cite{giorgini_theory_2008,liu_universal_2021}.

Quantum fluids of light provide a complementary limit in which nonlinear-wave hydrodynamics, vortices, and superfluid breakdown occur in optical media~\cite{boulier_microcavity_2020,jacquet_polariton_2020,lerario_vortex-stream_2020}. Some platforms, especially polariton fluids, are intrinsically driven-dissipative: pump, decay, reservoir coupling, and phase locking enter the basic equation of motion. Others are closer to conservative Gross-Pitaevskii dynamics; paraxial photon fluids in nonlinear media have been used to study nonlocal superfluid flow past obstacles~\cite{vocke_role_2016}, vortex-dipole conversion into Jones-Roberts rarefaction pulses~\cite{baker-rasooli_observation_2025}, and short-time dynamics of quantum turbulence~\cite{baker-rasooli_turbulent_2023,ferreira_exploring_2024}. Optical fluids offer opportunities for high precision measurements that are challenging in atomic gases~\cite{glorieux_chapter_2025} and hence a pathway to more rigorous tests of turbulent cascade phenomena in driven-dissipative steady states, and tests of coherent hydrodynamic processes in effectively non-dissipative propagation geometries.

\section{Outlook and open questions}

The frontier has now moved beyond observing disordered vortices, phonon populations, or familiar spectral exponents. Increasingly the focus is on establishing clearer evidence of phenomenology via links between the forcing scale, dissipation scale, inertial range, flux direction, incompressible/compressible spectral decomposition, vortex-resolved observables, wave-occupation spectra, and the finite-size, compressibility, and loss effects that shape real atomic gases. Cascade analysis in a finite quantum fluid needs power law evidence, but also knowledge of what is transported, over which range of $k$, via which flux, and into which dissipation channel.

Several open problems follow from this perspective. Experiments still need controlled steady inertial ranges in finite gases, direct flux measurements rather than cascade inference from exponents alone, and clean separation of vortex, wave, and mixed cascades. Theory must connect equilibrium vortex-gas thermodynamics to nonequilibrium turbulent transport without conflating Onsager clustering, decaying relaxation, and forced steady cascades. A further challenge is to understand how classical-looking structure functions and intermittency emerge from quantized circulation, vortex-core physics, compressible Gross-Pitaevskii dynamics, and, in real experiments, thermal damping and loss. These questions define the route from qualitative turbulence images to a quantitative understanding of atomic-gas turbulence.

The larger goal is to determine which features of turbulence are universal properties of coherent nonlinear quantum fluids, and which are platform-specific. Contact-interaction Bose gases provide the reference case where topological quantum turbulence, classical-field wave turbulence, compressibility, and their mixed regimes can be cleanly separated. Further richness is introduced in dipolar gases with anisotropic nonlocal forces and rotonic spectra; spinor and multicomponent gases add the possibility of coupling between external and internal turbulent degrees of freedom; fermionic superfluids add pair breaking and crossover physics. Experimental studies of exponents, fluxes, correlations, vortex statistics, and dissipation pathways across different regimes and systems may yield new insights into the universal aspects of superfluid turbulence and identify which of them depend specifically on quantum topology.

Quantum gases offer a view into the complex dynamics of a clean, dilute, and controllable system. Their high degree of controllability has allowed precise manipulation and observation of topological quantum-turbulent phenomena not accessible in classical fluids, with point-vortex clustering into negative-temperature Onsager states providing a prime example~\cite{gauthier_giant_2019,johnstone_evolution_2019}. Closely related studies of drag forces and vortex nucleation, both theoretically~\cite{frisch_transition_1992,winiecki_vortex_2000,reeves_identifying_2015,christenhusz_emergent_2025} and experimentally~\cite{raman_evidence_1999,onofrio_observation_2000,neely_observation_2010,kwon_critical_2015,kwon_observation_2016}, have provided evidence for universal aspects of drag in superfluids tied to the superfluid Reynolds number~\cite{onsager_introductory_1953,barenghi_is_2008,reeves_identifying_2015,christenhusz_emergent_2025}. The classical-field cascades realized in quantum gases, including the direct cascade~\cite{navon_emergence_2016}, provide controlled tests of Kolmogorov--Zakharov transport, while vortex-resolved experiments test how quantized circulation modifies turbulent relaxation and scale transfer. Together these regimes provide a route to separating universal transport phenomenology from effects that depend specifically on the quantum topology of the host fluid.

\section*{Acknowledgements}
This preprint will appear as a chapter in the Springer book entitled \emph{Short and Long
Range Quantum Atomic Platforms — Theoretical and Experimental Developments}
(provisional title), edited by P. G. Kevrekidis, C. L. Hung, and S. I. Mistakidis.

\makeatletter
\def\bibsection{\section*{\refname}\ifx\sectionmark\@gobble\else
    \markright{\refname}\fi
    \mtaddtocont{\protect\contentsline{mtchap}{\refname}{\thepage}\hyperhrefextend}%
    \csname biblst@rthook\endcsname\par}
\makeatother
\bibliographystyle{spphys}
\bibliography{turb_references}
%


%
%


\end{document}